\documentclass[11pt]{article}

\usepackage[utf8]{inputenc}
\usepackage[T1]{fontenc}
\usepackage{amsmath,amssymb,amsthm,bbm}
\usepackage{graphicx,transparent,eso-pic}
\usepackage[document]{ragged2e}
\usepackage{multicol}
\usepackage{url}
\usepackage[margin=2cm]{geometry}
\usepackage{booktabs}

\newcommand{\noun}[1]{\textsc{#1}}

\begin{document}

\title{Empowering Urban Governance through Urban Science: Multi-scale Dynamics of Urban Systems Worldwide}

\author{\noun{Juste Raimbault}$^{1,2,3,\ast}$, \noun{Eric Denis}$^3$ and \noun{Denise Pumain}$^3$\medskip\\
$^1$ CASA, University College London\\
$^2$ UPS CNRS 3611 ISC-PIF\\
$^3$ UMR CNRS 8504 G{\'e}ographie-cit{\'e}s\medskip\\
$^{\ast}$\texttt{juste.raimbault@polytechnique.edu}
}
\date{}

\maketitle

\justify

\begin{abstract}
The current science of cities can provide a useful foundation for future urban policies, provided that these proposals have been validated by correct observations of the diversity of situations in the world. However, international comparisons of the evolution of cities often produce uncertain results because national territorial frameworks are not always in strict correspondence with the dynamics of urban systems. We propose to provide various compositions of systems of cities to better take into account the dynamic networking of cities that go beyond regional and national territorial boundaries. Different models conceived for explaining city size and urban growth distributions enable to establish a correspondence between urban trajectories when observed at the level of cities and systems of cities. We test the validity and representativeness of several dynamic models of complex urban systems and their variations across regions of the world, at the macroscopic scale of systems of cities. The originality of the approach is in considering spatial interaction and evolutionary path dependence as major features in the general behavior of urban entities. The models studied include diverse and complementary processes, such as economic exchanges, diffusion of innovations and physical network flows. Complex systems’ dynamics is in principle unpredictable, but contextualizing it regarding demographic, income and resource components may help in minimizing the forecasting errors. We use among others a new unique source correlating population and built-up footprint at world scale: the Global Human Settlement built-up areas (GHS-BU). Following the methodology and results already obtained in the GeoDiverCity project~\cite{pumain2015multilevel,cura2017old,pumain2017urban}, including USA, Europe and BRICS countries, we complete them with this new dataset at world scale and different models. This research helps in further empirical testing to the hypotheses of the evolutionary theory of urban systems and partially revising them. We also suggest research directions towards the coupling of these models into a multi-scale model of urban growth.\medskip
\end{abstract}

\section{Introduction}

The urbanization process has profoundly transformed the distribution and organization of human societies on the surface of the earth since the emergence of the first cities some 10,000 years ago. Physically, it corresponds to a concentration of populations in densely populated built-up agglomerations, whose dimension, very unequal, expand over no less than four orders of magnitude (from $10^3$ to $10^7$ in number of inhabitants). Economically, urbanization translates into accumulations of capital, knowledge and wealth and the multiplication of networks that accompany creation and technological innovation. Socially, urbanization promotes the diversification and increasing sophistication of forms of institutional and political organization, including an intensification and refinement of the social division of labor. Culturally, urbanization, often held to be synonymous with ``civilization'', catalyzes the evolution of urban practices and collective representations through social mixing, hybridization and education, opening perspectives towards the possibility of a better knowledge, diversity and greater acceptance of the Other. All these trends seem to represent progress for the future of humanity, which is already reflected in synthetic indicators such as average income, the index of human development, or life expectancy in good health, despite a sharp increase in social inequalities since the 1980s and the neo-liberal shift in urban governance that accompanied the debt write-offs of cities and states. However, we know today that this evolution also marks the irruption of human activity in energy balances and the functioning of the terrestrial ecological system, which leads to propose a new stratum, called the Anthropocene, in geological time scales. The cities, which have accompanied and maintained the current overconsumption of the planet's resources, are part of a modification of human relationships with the Earth during this evolution that they seem threatened, not only in their future development, but sometimes even for their survival, by the scarcity of available materials and energy sources and the disasters that could be caused by major climate change initiated since at least two centuries. Is the success of urbanization responsible for that evolution and should the Anthropocene rather be called an Urbanocene?

Actually these cities, which have often shown great resilience in past centuries, may be a solution to the problem they could be a symptom of. Their organization, partially directed and partially spontaneous, in spatially distributed, hierarchical and complementary systems of cities for the exploitation, the control, but also the service, the maintenance and the adaptation of the territories, is undoubtedly a very important asset for a proper execution of the "ecological transition" being implemented in all parts of the world. This transition can succeed if it intelligently uses the evolutionary properties of city systems, ensuring both the "top down" dissemination of international regulations and the "bottom up" circulation of the many local initiatives in favor of technical and technological processes as well as organizations that will ensure respect for biological, cultural and geographical diversity, but also a more equal distribution of resources and wealth to reduce predatory relations to nature. Diversity in all these forms seems indeed essential to the continuation of the human and urban adventure. However, it is still necessary to develop a better understanding of this multi-scale urban dynamics before providing recommendations to planners and practitioners.  Geographers have collaborated with other specialists to build relevant knowledge of systems of cities, anchored in the knowledge and comparison of diverse world region’s contexts observed over many historical periods.

We illustrate in this paper how large sources of urban data and dynamic models when powerfully and safely handled with intense computing can help to identify the diversity of co-evolution regimes that have to be disentangled for being able to propose efficient solutions to the urban problems. More precisely, we tackle the question of how to understand the properties and dynamics of large systems of cities including complementary processes driving urban growth, using new sources of comprehensive urban data which are the Global Human Settlement Layer database~\cite{florczyk2019ghsl} and the Geodivercity database~\cite{pumain2015multilevel}. Our contribution relies on the following points: (i) we provide a theoretical framework to interpret evolutionary urban dynamics at the scale of systems of cities, building on the evolutionary urban theory proposed by~\cite{pumain2018evolutionary} and on the concept of co-evolution within urban systems defined by~\cite{raimbault2018caracterisation}; (ii) we study the empirical properties of large urban systems including patterns of urban growth and scaling properties, including different definitions of urban systems; (iii) we apply and calibrate simulation models for urban dynamics on 6 of these large systems worldwide, comparing very different processes including spatial interactions, transportation infrastructures, economic exchanges, and innovation diffusion, yielding for each urban system plausible underlying  mechanisms  driving their dynamic and providing potential policy insights.

The rest of this paper is organized as follows: in the next section, we develop the theoretical framework of the evolutionary urban theory and the underlying urban growth models based on spatial interactions; a third section complements this theoretical background by developing the concept of co-evolution within systems of cities. We then study empirical properties of large urban systems worldwide, and calibrate dynamical models of urban growth on these systems. We finally discuss future developments and possible implications for the sustainability of urban systems planning and management.

\section{Geographical models of urban growth within systems of cities}

The growth of cities is often interpreted according to the decisions taken locally by a multitude of very diverse stakeholders, for instance decisions about urban policies, locational strategies from firms, or motivations for residential migrations. These actions from urban stakeholders appear at a first sight as the direct causes of urban growth, at this micro-level of individual decisions. Obviously the growth or decline of a city is resulting from aggregating all these processes, and also from possible radical changes in environmental conditions.  But because of until recently a lack of data at this level of observation, most models of urban growth were developed at meso-geographical level for subsets of urban units. As cities mostly grow demographically and expand spatially by aggregating new populations and activities from their center toward peripheries according to monocentric or polycentric patterns, the growth should be computed for evolving urban agglomerations or functional urban areas that are properly delineated at each date of the period under observation. The comparability of many results about urban growth and distribution of city sizes is too often hampered because authors did not apply this geographical principle when defining and delineating properly the urban units they consider (for instance: \cite{eeckhout2004gibrat} or \cite{xu2014discontinuities}, see \cite{cottineau2017metazipf} for a full review). Systematic investigations were made recently after developing original harmonized data bases on thousands of urban agglomerations over decades and even centuries in the GeoDiverCity project~\cite{pumain2015multilevel,cura2017old,pumain2017urban}, including USA, Europe and BRICS countries.

Such consistent statistical observations on thousands of cities and over centuries, enabled at first to confirm the conclusion that emerged in pioneer works on the growth process within integrated systems of cities (i.e. well connected cities obeying a unified system of political, cultural and economic rules that currently define a ``territory''): each city has a probability of growing similar to other cities belonging to the same territorial system. This was characterized as a `` distributed growth'' process that can be observed on the long run with many local and temporal fluctuations in any system of cities all over the world \cite{gibrat1931ingalits,robson1973view,pumain1982dynamique}: Gibrat’s model of urban growth constitutes a first good approximation of the distribution of urban growth within a system of cities. Gibrat's ``law of proportional effect'' means that growth rates are equiprobable whatever city size and are not correlated with previous growth rates.

This rather good fit already provides a double explanatory gain: it explains the remarkable persistence of urban spatial patterns and hierarchies over very long periods of time that may exhibit such meta-stability with very little changes over centuries. And moreover it provides an ``explanation'' for the statistical shape of urban sizes distribution (that is lognormal, close to Zipf’s law or other types of skewed distributions), according to a stochastic process that was also anticipated by Herbert Simon in 1955. Conversely, despite attempts at developing an economic interpretation of the genesis of urban hierarchies through firm choices or preferential attachment~\cite{gabaix1999zipf}. P. Krugman still considers it as ``a mystery'' regarding its explanation within economic theory~\cite{krugman1996confronting}.

Many observed distributions of city sizes (actually: settlement sizes including hamlets, villages, towns and SMAs) are close to lognormal distributions (evidence from \cite{robson1973view,pumain1982dynamique,eeckhout2004gibrat,decker2007global} and Gibrat's growth model mathematically leads on the very long run to a lognormal distribution of city sizes, but some of the hypothesis of Gibrat's growth model are sometimes partially rejected through statistical testing: more or less high correlations may be found between growth rates and city size (most of the time positive correlation), and positive correlation between successive growth rates also may be observed for some time periods.

That is why complementary models have been proposed that offer better fits when adding the possible effects of spatial interactions to Gibrat's model (which is a stochastic model where urban agglomerations are represented as independent units, which of course cannot be a relevant geographical model for epistemological reasons). \cite{favaro2007croissance} and \cite{favaro2011gibrat} developed a geographical model of urban growth including: a recurring emergence of clustered new innovations that create growth cycles (following a Schumpeterian process); a spatial diffusion of innovations occurring through a dynamic spatial interaction model \cite{wilson1971family}; a spatial diffusion of innovation according to a hierarchical process \cite{hagerstrand1952propagation}. The growth of a city depends on its share of labour force in each innovation cycle that induces a scaling parameter larger than one for the urban activities that participate in the innovation wave of the moment.

Analytically, the model can be expressed in a form that is very close to Gibrat’s model:

\begin{equation}
    \log{P_{i,t}} = \alpha + \log{P_{i,t-1}} + G_i + u_{i,t}
\end{equation}

where $G_i$ holds for the ``bias'' noticed in estimating Gibrat's model by Ordinary Least Squares (linked with spatial interaction processes, see \cite{favaro2011gibrat}). The noises $u_{i,t}$ are stochastic variables in time and space with zero averages, which allow capturing random fluctuations. They correspond to the random distribution of growth rate in the initial Gibrat's model. The endogenous growth rate $\alpha$ and the growth induced by interactions are deterministic given a realization of populations at the previous step.

This model has the advantage to replace a generic statistical model of growing independent entities (Gibrat’s urban growth model) by \emph{a model of spatially and temporally interdependent entities} (i.e. the geographical concept of ``system of cities'' or ``settlement system''). It reproduces the observations on differential scaling parameters for urban activities according to their age in innovation cycles~\cite{pumain2006evolutionary}. It also makes explicit the multilevel dynamics of interurban competition for capturing innovation, which may itself generate new innovation through interurban emulation, within an evolutionary perspective. 

Moreover, this model enables us to interpret variants that represent the path dependence effects occurring through territorial differentiation. For instance, in new urban systems, as in USA, there is a spatial filling process that occur through spatial waves of urban growth (urban Frontier) corresponding to the diffusion of economic cycles \cite{bretagnolle2007formes}, whereas in mature urban systems, as in France, the innovations diffusion reaches cities that are not spatially regularly arranged but already experienced other growth periods according to distinct cycles of urban specialization~\cite{paulus2004coevolution,finance2016villes}.

Other types of models based on interactions have been introduced within the framework of the evolutionary urban theory~\cite{pumain2006evolutionary}. A deterministic version of the previous Favaro-Pumain model, based on economic exchanges between cities, was introduced by \cite{cottineau2014evolution}. This model called Marius was originally designed for the Former Soviet Union but applies to any system of cities, as only population trajectories are required to be parametrized to estimate and evaluate the model. Scaling laws are used to extrapolate economic activities of cities based on population, and economic exchanges are then simulated. The model is more precisely a \emph{family of models}, since different processes can be taken into account in a multi-modeling way (for example, local impact of energy resources, switch between top-down planned economic structure and bottom-up interactions only). \cite{cottineau2015growing} formulated this methodological framework as an ``evaluation-based incremental modeling method'', which allows testing of concurrent hypotheses to explain trajectories of systems of cities.

This type of models was also applied to the subject of transportation networks within urban systems. The SimpopNet model introduced by \cite{schmitt2014modelisation} captures entangled trajectories of city populations and network links between these cities in a stylized setting. It was shown by \cite{raimbault2020unveiling} that it effectively reproduces co-evolutionary dynamics (in a precise sense we develop below), which is a central theoretical feature in evolutionary urban theory. More recently, \cite{raimbault2020indirect} proposed a simplified version of the Favaro-Pumain model, in which the deviation due to interactions includes physical network effects, through direct spatial interaction models, but also indirect feedback of network flows on city growth. The model was calibrated on real data for France over a long time span (1830-2000). It was extended to a co-evolutionary model in \cite{raimbault2018modeling}, also calibrated on French data with railway network accessibility, for which the capacity to capture a broad range of co-evolutionary regimes was demonstrated.

These different models have the particularity to enter in the same theoretical frame of the evolutionary urban theory, of being backed-up by empirical studies and datasets constructed during the GeoDiverCity ERC project \cite{pumain2015multilevel}, and to simulate urban systems population trajectories on long time scales and large spatial scales (macroscopic scale of a region, country or continent). They are a fundamental component of our evidence-based approach to urban systems, and will be applied and compared in a benchmark below.

\section{Systems of cities and co-evolution regimes}

We first need to introduce another theoretical component justifying the use and comparison of these different models within the evolutionary urban theory. Evolutionary economics~\cite{nelson2009evolutionary} have a wide experience in transferring the concept of evolution from its initial context in biology \cite{durham1991coevolution}. This requires to be able to identify within the studied systems several components, which are transmission processes, transformation processes and differentiation between populations of agents emerging from these processes. Artificial Life literally studies such artificial systems, extending biological systems to ``life as it could be'' \cite{langton1997artificial,bedau2003artificial}. Although the core component are not always explicit, and these concepts are sometimes relaxed to long-time structural dynamics of systems components, evolutionary concepts have been successfully applied in economics as mentioned above, but also in social sciences with the study of cultural evolution~\cite{mesoudi2017pursuing} or the evolution of social organizations \cite{volberda2003co}. The interplay between social components and biological components can even be considered \cite{bull2000meme}.

In that context, co-evolution originally corresponds to evolutionary changes in two species in strong mutual influence \cite{janzen1980coevolution}. The idea of diffuse co-evolution, taking into account the environment in which co-evolution takes places, including a broad network of other species, was proposed as a refinement of this concept \cite{strauss2005toward}. To effectively designate co-evolutionary processes, or more generally strongly entangled dynamics, it was also developed in economic geography \cite{schamp2010notion}. Thematic applications include the study of economic clusters \cite{ter2011co}, of technological change \cite{colletis2010co}, or environmental economics \cite{kallis2007coevolution}.

In geography, it was particularly developed by the evolutionary urban theory~\cite{pumain1997pour}, which in practice consists in a dynamical non-equilibrium approach to urban systems as complex adaptive systems, in which interactions between components are central \cite{paulus2004coevolution,schmitt2014modelisation}. In that context, building on the definition introduced by~\cite{raimbault2018co,raimbault2018caracterisation}, we propose a multi-level definition of co-evolution which is consistent with the evolutionary urban theory~\cite{pumain2018evolutionary}. First of all, transformation processes of territorial components should induce evolutionary dynamics at different scales. It is not clear what would be ``an urban genome'', but empirical evidence of innovation diffusion and cultural evolution (including social progress and changes in governance), but also physical transformations or physical flows between cities, suggest that these dynamics can be interpreted as evolutionary in a loose sense. Then, co-evolution can occur at a microscopic level between particular artifacts or agents (coupled dynamics), at a mesoscopic scale of a population of entities (in the initial biological and statistical sense, i.e. that characteristics of two populations of entities mutually influence themselves in their evolutionary trajectories), and finally at a macroscopic scale in the usual sense used in geography as strongly entangled dynamics at the system level.

This definition implies the existence of co-evolution niches in the sense of \cite{holland2012signals}, which can be understood as a system of boundaries and corresponding subsystems in which co-evolution takes place. This pertains particularly well for territorial systems, for which the niches will be consistent territorial ensembles, imbricated at different scales. The cities at a regional level are for example a first level of niche, embedded into the urban system at a larger territorial scale as the national, continental and even global one. Within a given niche, the co-evolving populations of entities will be in a specific co-evolution regime, concept introduced by \cite{raimbault2017identification} in the case of transportation networks and territories, to describe a given causal network between the variables considered. For example, population of cities and their centrality in the transportation network may be in a circular causal relationship (taken as a weak Granger causality), or in a unidirectional relation, or even in a triangle relationship with a third variable such as accessibility. Each case is a particular co-evolutionary structure - even if strictly speaking there is no co-evolution when there is no circularity, the term of co-evolution regime can be used in a broad way to describe this state of dynamical relationships.

This directly implies that (i) urban systems worldwide have each their own co-evolutionary trajectory, and thus their own driving processes; (ii) the way they could merge into a global urban system implies possibly other types of processes (emergence of mega-city regions e.g.) and is not incompatible with the persistence of local regional urban systems. A quantitative investigation of the second point is still out of the scope of this preliminary work, as it would imply more complicated and possibly multi-scale models, but also more accurate global data on a longer time span. We will however provide a first experiment below by testing models on a new global database. The first point is consistent with the idea of comparing different models with underlying processes of a different nature to try to reproduce the trajectories of urban systems, and this for different systems across the world. We will also provide results for such experiments below. The test of such dynamical models is an indirect way to dig into the diversity of co-evolution regimes, as data is generally missing for a direct investigation. The model fitting the best a given urban system and corresponding extrapolated parameters (for example interaction distance and hierarchy for a spatial interaction model) provide information on the underlying co-evolution dynamics, and can even in some case be associated with regimes directly identified in synthetic settings (such as the ones found by \cite{raimbault2018modeling}, but this elaborated investigation also remains out of the scope of this paper).

Building on this theoretical background, we will now investigate a new global database of urban areas and some stylized facts that can be extracted from it, and then test and compare dynamical models for urban systems.

\section{A new source of data for comparing urban trajectories worldwide}

We test here the generality and robustness of stylised facts on systems of cities when using a new source of comparable urban data provided by the Global Human Settlement layer dataset (GHS\footnote{Source: GHSL (Global Human Settlement Layer) : GEO Human Planet Initiative (European Commission)}). The dataset was already explored statistically for comparing trends of urban sprawl in the countries of the world by \cite{denis2020population}, while \cite{melchiorri2019principles} computing a land use efficiency index measured a global trend towards densification with a diversity of urban trajectories according to regions of the world. The data delineate comparable morphological urban agglomerations by detecting built up area from satellite images at 40 m resolution, and combining it with a population layer generated at 250m resolution using local data at municipal and district levels provided by censuses on a regular 1 km\textsuperscript{2} grid \cite{dijkstra2014harmonised}. 

The dataset is available at different dates between 1975 and 2015. In 2015 the source delineated precisely some 13,000 urban agglomerations larger than 50,000 inhabitants in the world. For each urban agglomeration figures on built-up surface and population are provided in 1975, 1990, 2000 and 2015. It also includes measurements of GDP, green areas and pollution levels from 1990 to 2015. 

The short period of time considered is the major limitation of the source, that can be mitigated because of the multiple case studies that are now available from comparative analysis of urban trajectories over much longer time spans, such as \cite{robson1973view} for England and Wales 1801-1960; \cite{pumain1982dynamique} for France 1831-1975; \cite{bretagnolle1999systemes} for France 1831-1990; \cite{bretagnolle2007formes} for the US 1800-2000; \cite{swerts2013systemes} for India 1901-2011 and China 1982-2010; \cite{rozenblat2018international} on many large regions in the world and various time-spans.

\cite{dijkstra2018applying} have compared the GHSL data with the statistics provided by the World Urbanization Prospects \cite{desa2018world} and found a rather good fit with the data on individual cities larger than 300,000 inhabitants. We confirm that the GSHL source provides statistics that are roughly compatible with the results that were obtained in different countries of the world using our dedicated harmonized data bases (Table~\ref{tab:populations}). The total urban populations of each country or region should be smaller in GHSL because in GeoDiverCity we considered urban agglomerations larger than 10,000 inhabitants, but this is not always the case, especially for India and China whose urban populations could be overrepresented in GHSL. However, computing the slope of rank size distribution brings comparable results about the different structures of urban hierarchies, with less unequal distributions of city sizes in countries that developed earlier their systems of cities \cite{pumain2015multilevel}.

\begin{table}
\caption{\textbf{Representativeness of GSHL source compared to harmonized data bases from GeoDiverCity.} Population is given in Millions. All indicators are given for the GHSL database, except for the population from the Geodivercity database for comparison (third column). The Rank-size exponent is estimated with a standard OLS on logarithms. Standard deviations of the rank-size exponent are all smaller than 0.02, and the adjusted R-squared larger than 0.97. FSU = Former Soviet Union.\label{tab:populations}}
\centering
\begin{tabular}{cccccc}
    \toprule
        System & Pop (M) & Pop geodiv. & Cities & Rank-size  \\
        \midrule
         Europe & 188 & 291 & 693 & $0.94$  \\
         China & 567 & 481 & 1850 & $0.91$  \\
         Brazil & 112 & 161 & 349 & $0.99$  \\
         India & 703 & 427 & 3248 & $0.78$  \\
         South Africa & 25 & 25 & 77 & $1.05$ \\
         US & 153 & 324 & 287 & $1.16$  \\
         FSU & 120 & 174 & 450 & $0.92$ \\
         \bottomrule
\end{tabular}
\end{table}

In order to roughly check the global reliability of the data set we computed the now classical indices describing urban hierarchies with the slope of Zipf's rank-size rule. This can be done for a diversity of ways of grouping cities, i.e. dividing the world in consistent systems of cities. Delineating proper systems of cities at the macro-level of inquiry is an even more delicate exercise than delineating urban entities at meso-geographical level. Theoretically one should consider subsystems that have more internal than external interactions with other cities, which raise difficult issues of determining which interactions to consider and over which period of time. Even if it was possible to identify and measure well adapted data, in most cases we would find that the largest cities of any system of cities have a much larger scope of interactions than the smaller towns \cite{bretagnolle2010simulating}. We shall try to elaborate more on this particular problem in a next paper, thus in this first attempt here we choose to experiment on different types of groupings, using the seven national or regional urban systems for which we had developed alternative databases in the GeoDiverCity programme \cite{pumain2015multilevel} and comparing with two other ways of grouping countries: a first one according to the main five continents of the world as many studies on flows of air flight passengers or even networks of branches owned by multinational firms demonstrated that they often encompass clusters of stronger internal interactions and have discontinuities between them; a second one is made according to major economic trade zones that are also supposed to be subsets of denser inter-urban exchanges. 

\begin{figure}
\centering
\includegraphics[width=\textwidth]{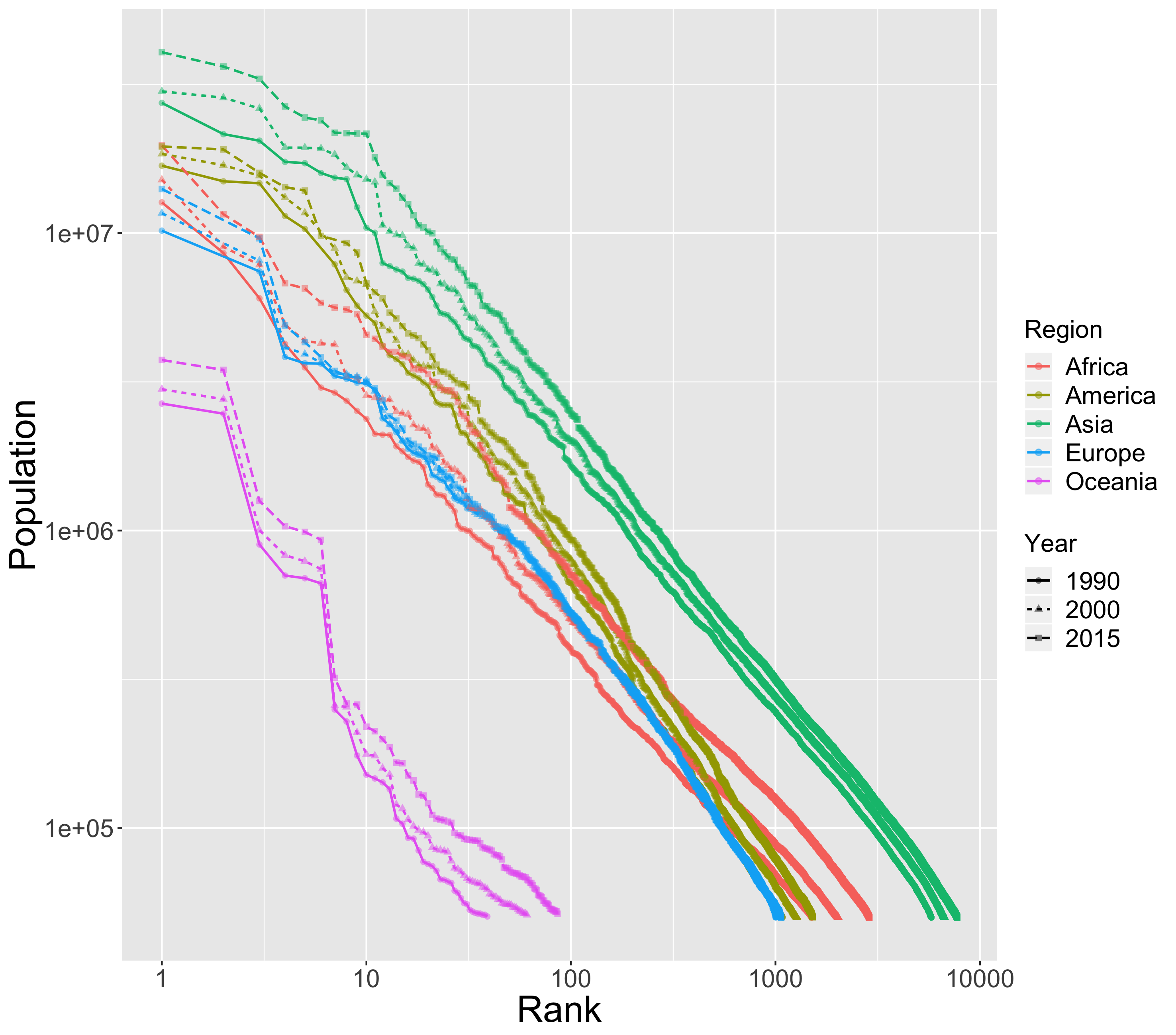}
\caption{Evolution of rank-size distributions by continents 1990-2000-2015.\label{fig:fig1}}
\end{figure}

A first simple statistical analysis confirms results on stylised facts that were already observed on many systems of cities. We show in Figure~\ref{fig:fig1} the rank-size plot for all 5 continents, at three different dates. We find as already known that the urban hierarchy roughly remains constant in time. Numerical values for exponents in this estimation are provided in Table~\ref{tab:tab2} for the year 2015. We also retrieve the fact that more mature urban systems (Europe) have a exponent closer to one, while recently booming urban systems such as Africa have an exponent far from one (0.78 in 2015) – such a result may also be linked with the absence of smaller towns in GHSL. Asia has an intermediate value of 0.87, what would be consistent with the fact of mixing subsystems which have a very long history (Japan and China e.g.) but also recently underwent drastic demographic transitions, with other subsystems whose development is more rapid and recent (South-East Asia). The primacy indices are also consistent with what could be expected: note that it is larger for Europe when Moscow is included (it would be close to one if taking the EU only for example, Paris and London being approximately the same size).

\begin{table}
\caption{Rank-size by continents in 2015. Source: GHSL, European Commission.\label{tab:tab2}}
\centering
\begin{tabular}{cccccc}
\toprule
System & Population (M) & NB of cities & P1/P2 (primacy) & Slope of rank-size & R2 \\
\midrule
Europa & 288 & 1067 & 1.45 & $0.93 \pm 0.003$ & 0.991 \\
America & 547 & 1521 & 1.02 & $1.02 \pm 0.002$ & 0.996 \\
Asia & 2143 & 7737 & 1.12 & $0.87 \pm 0.0004$ & 0.998 \\
Africa & 585 & 2876 & 1.7 & $0.78 \pm 0.0008$ & 0.997 \\
Oceania & 19 & 86 & 1.08 & $0.91 \pm 0.027$ & 0.926\\
\bottomrule
\end{tabular}
\end{table}

\begin{figure}
\centering
\includegraphics[width=\textwidth]{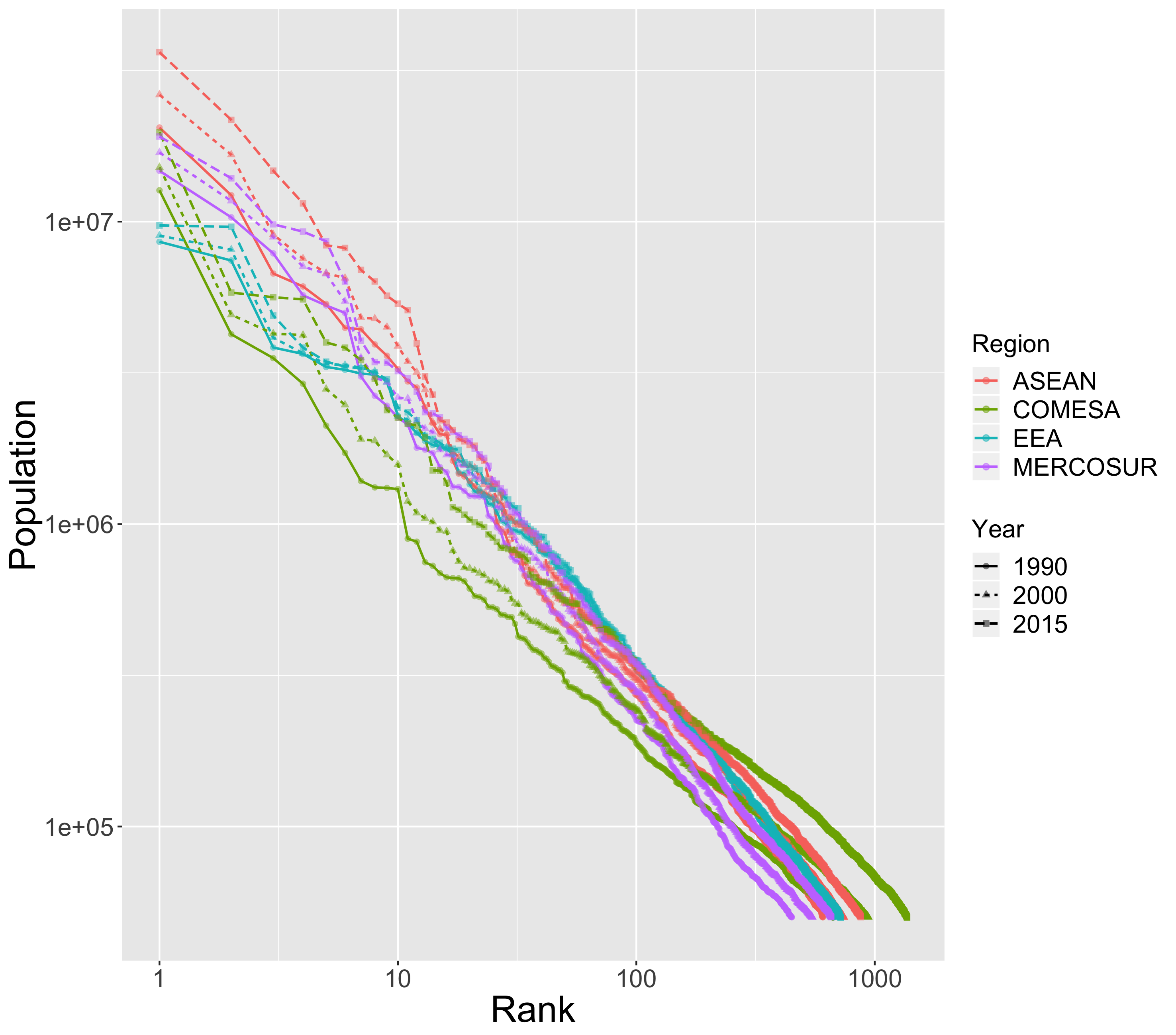}
\caption{Evolution of rank-size distributions by trade zones 1990-2000-2015.\label{fig:fig2}}
\end{figure}

We also study the rank-size properties by grouping the countries into trade areas, which are supposed to capture subsystems with a higher level of interurban interactions. We show the temporal evolution of rank-size plot in Figure~\ref{fig:fig2}, and corresponding statistics in 2015 in Table~\ref{tab:tab3}. Although the trade areas considered highly overlap with continents (in particular for EEA with Europe and ASEAN with Asia), the exponents obtained are different from the previous ones, and closer to one. According to \cite{arcaute2020scaling}, varying the definition of cities yield varying Zipf exponents, with no endogenous privileged definition. We can indeed expect the same when varying the system of cities considered, as \cite{corral2020truncated} show that power-law are not valid anymore when considering the tail of the distribution, i.e. changing the number of cities included in the estimation. If Zipf’s law was a pure product of ergodic stochastic processes without any interactions between cities, randomly sampling a subset of a given system should yield roughly the same exponent (at least with large sample size and with an OLS estimator). Although the sampling is not random here (a more elaborated statistical analysis remains for future work), the discrepancy suggests the opposite of the previous case and therefore that urban systems are highly non-ergodic and that interactions between cities are crucial. We suggest that the exact same phenomenon occurs for urban scaling laws, since they are in the same way strongly dependent to system definition \cite{arcaute2015constructing,cottineau2017diverse}, and that the claims of universality by the mainstream scaling literature are at least not compatible with this empirical evidence, at worse inaccurate in terms of underlying dynamical processes.

\begin{table}
\caption{Rank size for trade zones in 2015. Source: GHSL, European Commission. [ASEAN =Association of Southeast Asian Nations, 10 countries; MERCOSUR= Southern Common Market, 4 countries; COMESA= Common Market for Eastern and Southern Africa, 21 countries; EEA= European Economic Area, 31 countries]\label{tab:tab3}}
\centering
\begin{tabular}{cccccc}
\toprule
System & Population (M) & NB of cities & P1/P2 (primacy) & Slope of rank-size & R2 \\
\midrule
ASEAN & 293 & 874 & 1.67 & $0.92 \pm 0.003$ & 0.993 \\
MERCOSUR & 220 & 657 & 1.37 & $1.00 \pm 0.0016$ & 0.998 \\
COMESA & 252 & 1367 & 3.39 & $0.72 \pm 0.0014$ & 0.995 \\
EEA & 194 & 720 & 1.01 & $0.94 \pm 0.0026$ & 0.994 \\ 
\bottomrule
\end{tabular}
\end{table}

\begin{figure}
\centering
\includegraphics[width=\textwidth]{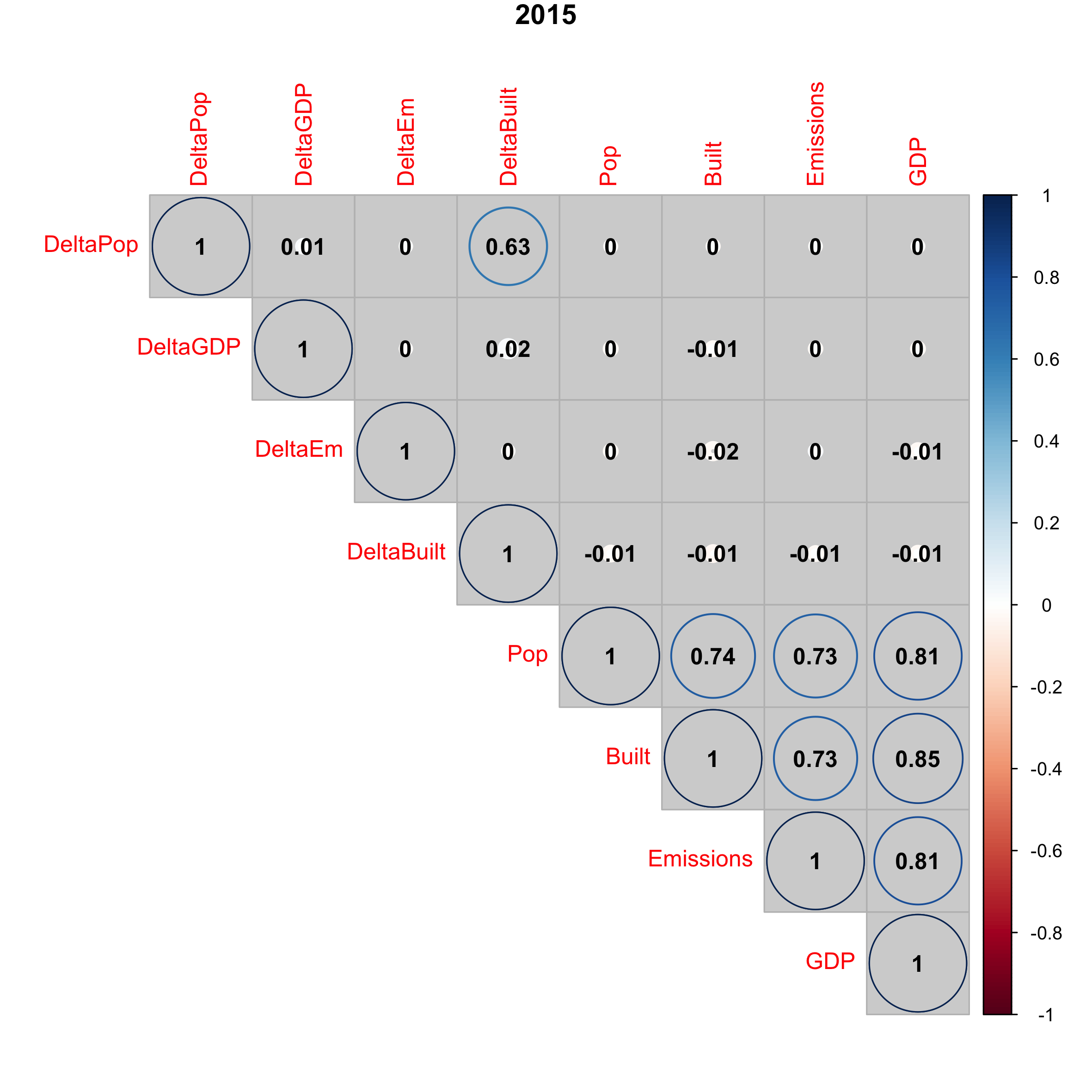}
\caption{Correlations between GHSL indicators (GDP, Emissions, Built-up surfaces, Population  and their evolution 2000-2015 (Delta)).\label{fig:fig3}}
\end{figure}

\begin{figure}
\centering
\includegraphics[width=\textwidth]{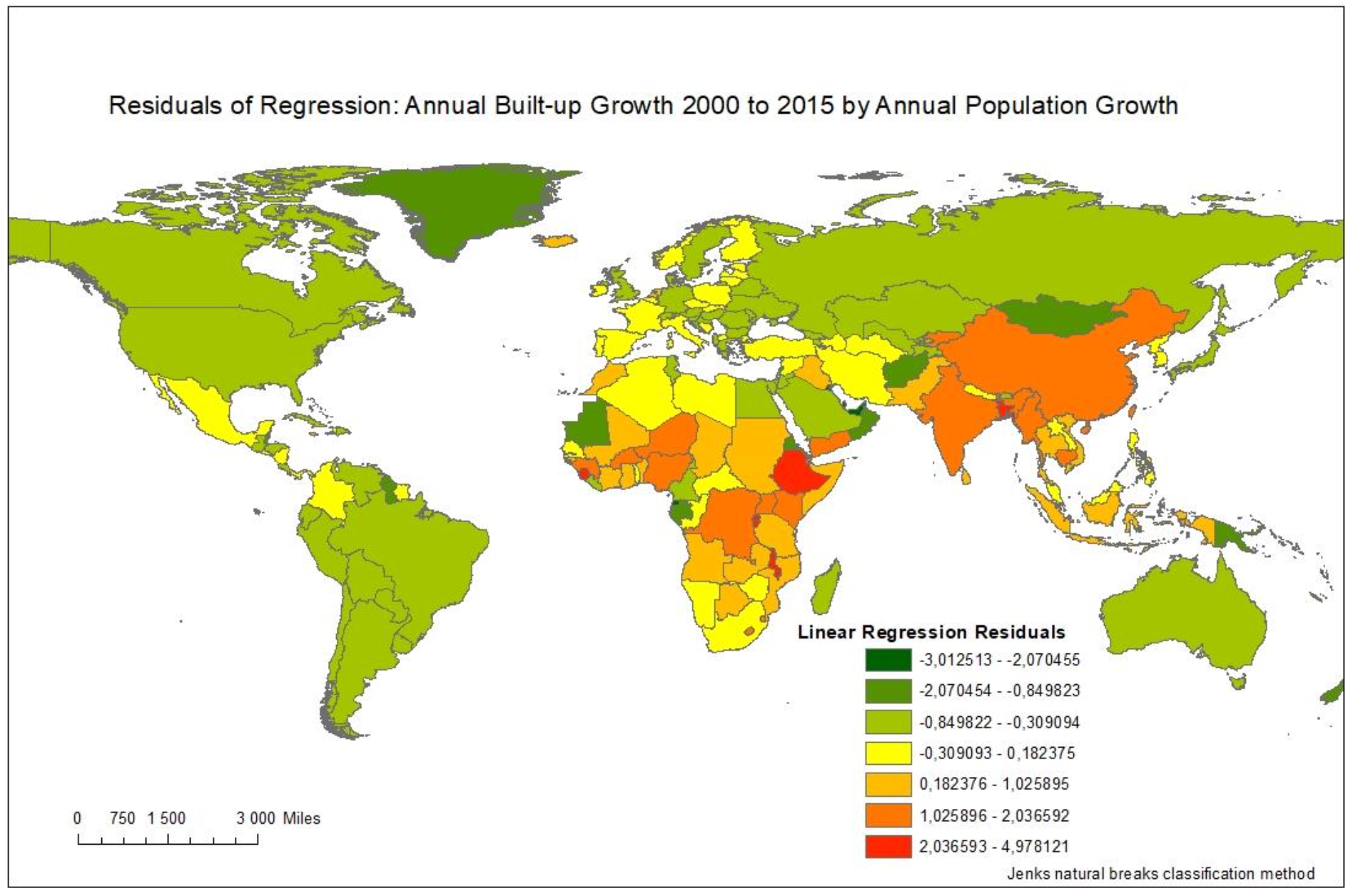}
\caption{Comparing spatial and demographic urban expansion 2000-2015. Source: GHSL, European Commission.\label{fig:fig4}}
\end{figure}

We also proceed to a simple statistical analysis of other variables in the database. We confirm the basic assumption of the Gibrat's model that population and population growth are uncorrelated. The correlation coefficients on Figure~\ref{fig:fig3} are above 0.7 between built-up area and population, as well as GDP and pollutant emissions. Cities are places of concentration of human activities with their desirable and less desirable outputs. However, there are wide inequalities in urban densities according to the continents (Asian cities are more than twice as dense as European cities that are ten times denser than North American and Australian ones, cf \cite{bertaud2003spatial}). Even if the progression of built-up area between 2000 and 2015 is positively correlated with the one of population, with a 0.63 coefficient, there are discrepancies between regions of the world in the evolution towards rather compact or more spread urbanization (Figure~\ref{fig:fig4}). The map shows clearly that sub-Saharan regions in Africa and south and southeastern Asia are expanding more rapidly the urbanized surfaces than their urban population – although being as well regions of rapid demographic urban growth. The variations of other variables, i.e. CO2 emissions and GDP, are totally uncorrelated with the variations of population, what means that their recent dynamics are independent.

We also compute the exponents of the data on GDP and emissions as power functions of urban sizes, under the hypothesis that they would exhibit scaling laws. Actually, the relationship between urban indicators in GHSL are not linear but follow scaling laws with exponents above 1, indicating a larger production and pollution potential when considering cities of larger size. This can be observed for subsets of cities we had previously observed in the GeoDiverCity programme in Table~\ref{tab:tab4}. The adjustments are in several cases of a low quality (for India for example), suggesting the relative validity of scaling laws. We did not proceed to additional tests to verify if these relations are effectively power laws (see \cite{broido2019scale} for the case of scale-free networks), but the chances are high for not having power-laws in a strict statistical sense. The distributions are still loosely fat-tailed and the exponents provide useful information. They lie close to one in the case of built-up area as expected, supra-linear but close to one for GDP also as expected, and supra-linear with a high exponent for emissions. The last relation on emissions was less expected, since according to \cite{ribeiro2019effects} the literature has very heterogeneous conclusions on emissions scaling, and the results vary highly depending on city definitions. There is however a high supra-linear scaling when considering urban areas, which is consistent with our findings. We obtain similar order of magnitudes (maximal variations of 0.2) when considering other geographical areas as above (continents and trade areas), confirming a certain robustness in the qualitative interpretations that can be done from these estimations. Note that this database holds promises for future more robust and useful scaling studies, as it has a worldwide coverage and the systems considered can be varied at ease, and includes fundamental variables related to urban issues (many scaling studies focusing on anecdotal variables such as the number of cinemas).

\begin{table}
\caption{Scaling exponents for large urban systems in 2015. We provide for each system the OLS estimates on logarithms of each variable as a function of population, with standard errors and adjusted R squared. FSU= Former Soviet Union. Source: GHSL, European Commission.\label{tab:tab4}}
\centering
\begin{tabular}{cccccc}
    \toprule
    System & Built-up area & GDP & Emissions \\
    \midrule
Europe & $0.93 \pm 0.016$ (0.83) & $1.15 \pm 0.019$ (0.83) & $1.50 \pm 0.038$ (0.69) \\
China & $1.06 \pm 0.019$ (0.62)  & $1.14 \pm 0.011$ (0.85)  & $1.84 \pm 0.037$ (0.57) \\
Brazil & $0.98 \pm 0.025$ (0.81)  & $1.10 \pm 0.055 $ (0.54) & $1.71 \pm 0.053$ (0.75) \\
India & $1.34 \pm 0.031$ (0.36) & $1.25 \pm 0.022$ (0.50) & $1.54 \pm 0.031$ (0.42) \\
S. Africa & $1.18 \pm 0.090$ (0.69) & $1.08 \pm 0.028$ (0.95) & $1.56 \pm 0.087$  (0.81) \\
US & $0.97 \pm 0.015$ (0.92) & $1.04 \pm 0.069$ (0.99) &$ 1.34 \pm 0.03$ (0.84) \\
FSU & $0.97 \pm 0.035$ (0.63) & $1.17 \pm 0.041$ (0.65) & $1.95 \pm 0.088$ (0.52)\\
\bottomrule
\end{tabular}
\end{table}

\section{Comparing dynamic models of urban growth}

We now turn to the test of different dynamical models to simulate the observed population trajectories of cities. As detailed above, this should allow to compare the influence of different possible drivers of urban growth, and possibly investigate the diversity of co-evolution regimes. We apply four different families of models, on the GHSL database between 1990 and 2015 and on the GeoDiverCity database. The first one is the stochastic model of Gibrat that can be used as a benchmark because it has the smallest number of hypotheses, without any interactions between cities. The second model inspired from \cite{favaro2007croissance} introduces innovation waves as impulses of urban growth with hierarchical diffusion, as detailed in section 2. A third model family is the Marius model, introduced by \cite{cottineau2014evolution}, which focuses on economic interactions between cities. The last model family is a spatial interaction model taking into account physical networks, studied by \cite{raimbault2020indirect}. The underlying processes within each of these model families are potential partial explanations for urban growth, and each belong to very different classes of processes.

\begin{figure}
\centering
\includegraphics[width=\textwidth]{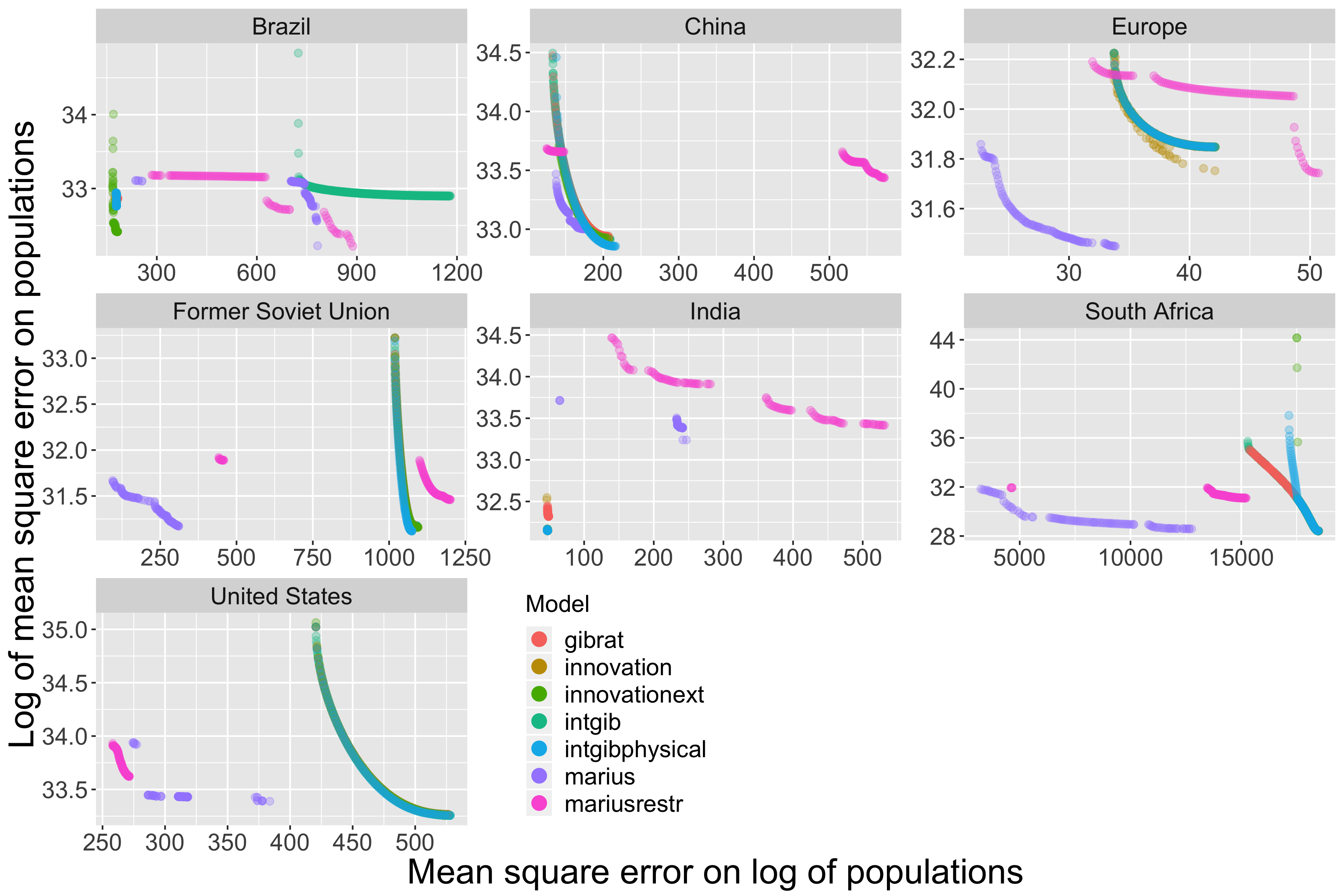}
\caption{\textbf{Test of six dynamic models simulating urban trajectories on systems of cities at national scale.} \textit{(Legend common to figures 5 to 9)}: two measurements of the fitness of each dynamic model (log of sum of deviations between observed and simulated populations on y axis [ensuring a better fit of the top of urban hierarchies] and sum of deviations between the logarithms of observed and simulated population of each city [ensuring a better fit for medium and smaller towns]) are related on the Cartesian graph exhibiting Pareto fronts that can be compared for the seven dynamic models. The calibration of exploration models were done using genetic algorithms and computation grid with the OpenMOLE platform.\label{fig:fig5}}
\end{figure}

\begin{figure}
\centering
\includegraphics[width=\textwidth]{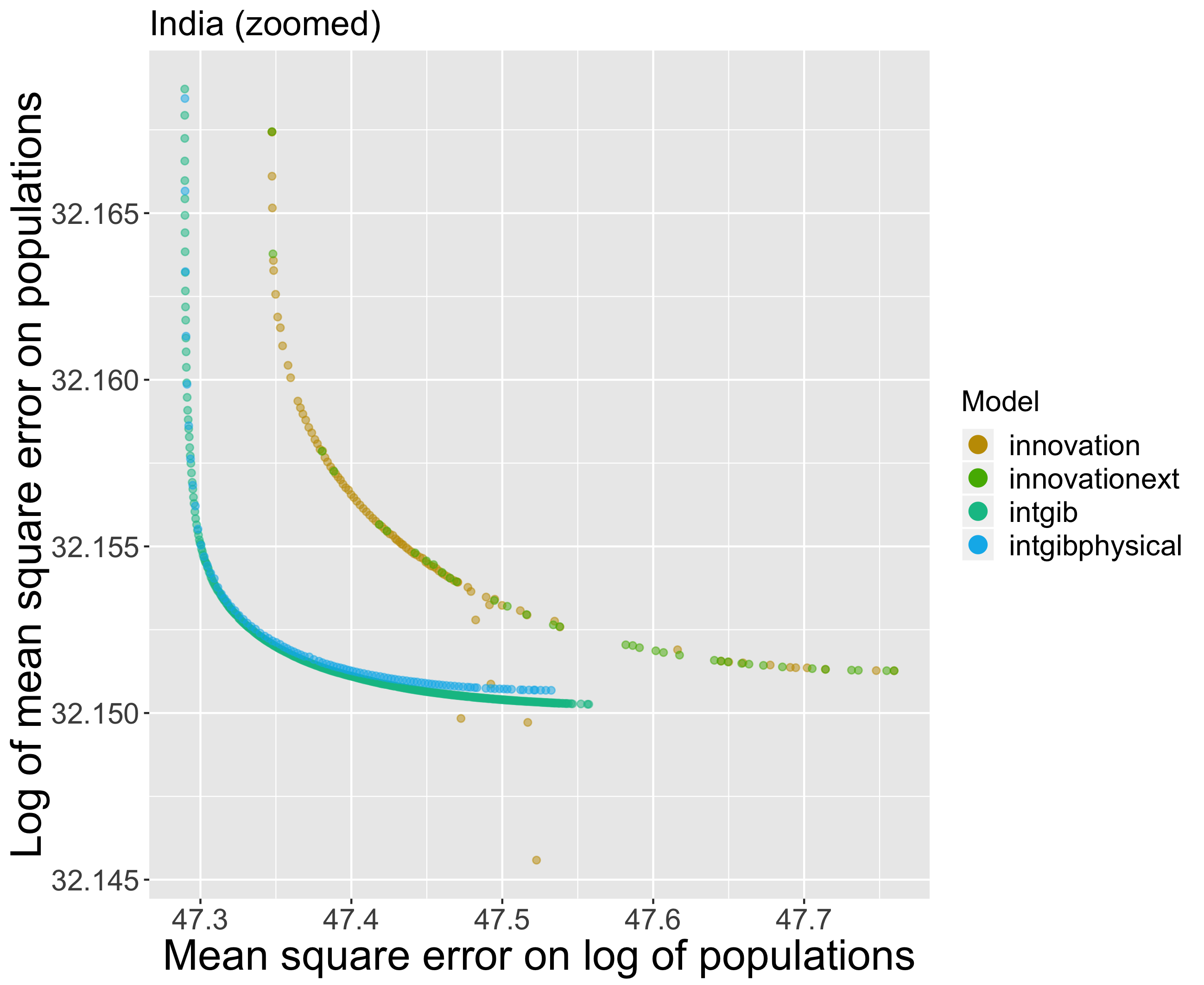}
\caption{Testing urban dynamic models on the Indian case.\label{fig:fig6}}
\end{figure}

\begin{figure}
\centering
\includegraphics[width=\textwidth]{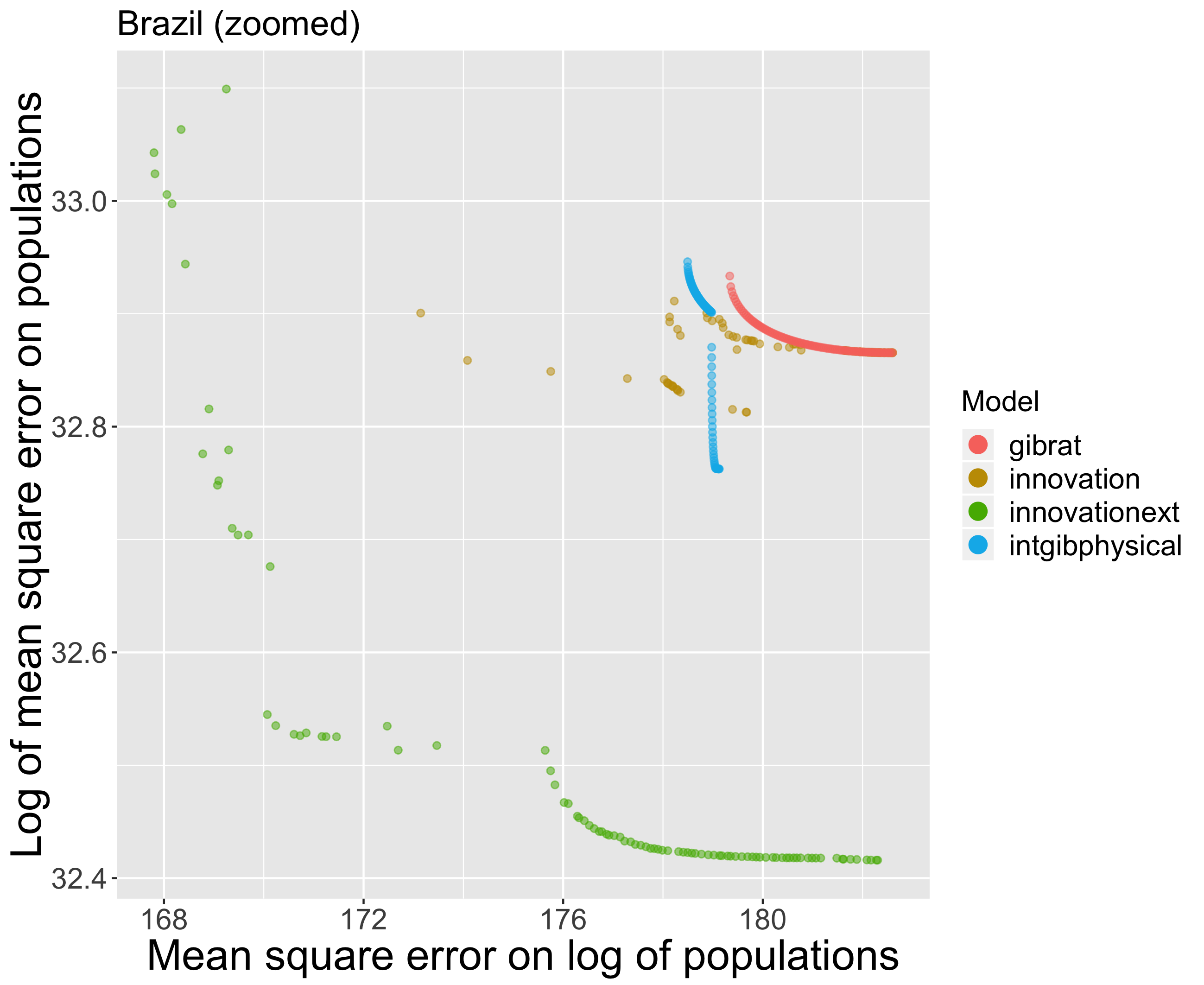}
\caption{Testing urban dynamic models on the Brazilian case.\label{fig:fig7}}
\end{figure}

\begin{figure}
\centering
\includegraphics[width=\textwidth]{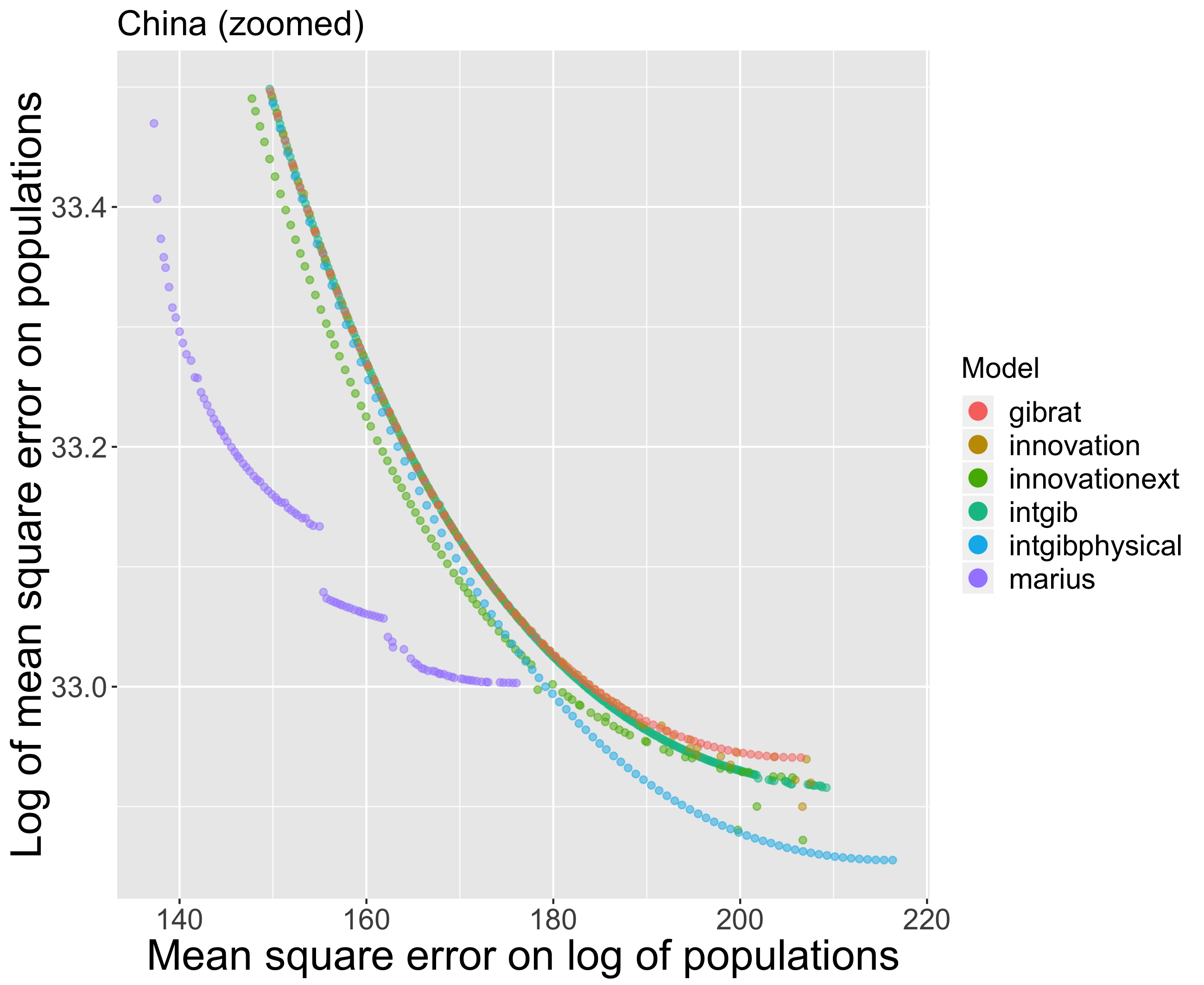}
\caption{Testing urban dynamic models on China.\label{fig:fig8}}
\end{figure}

\begin{figure}
\centering
\includegraphics[width=\textwidth]{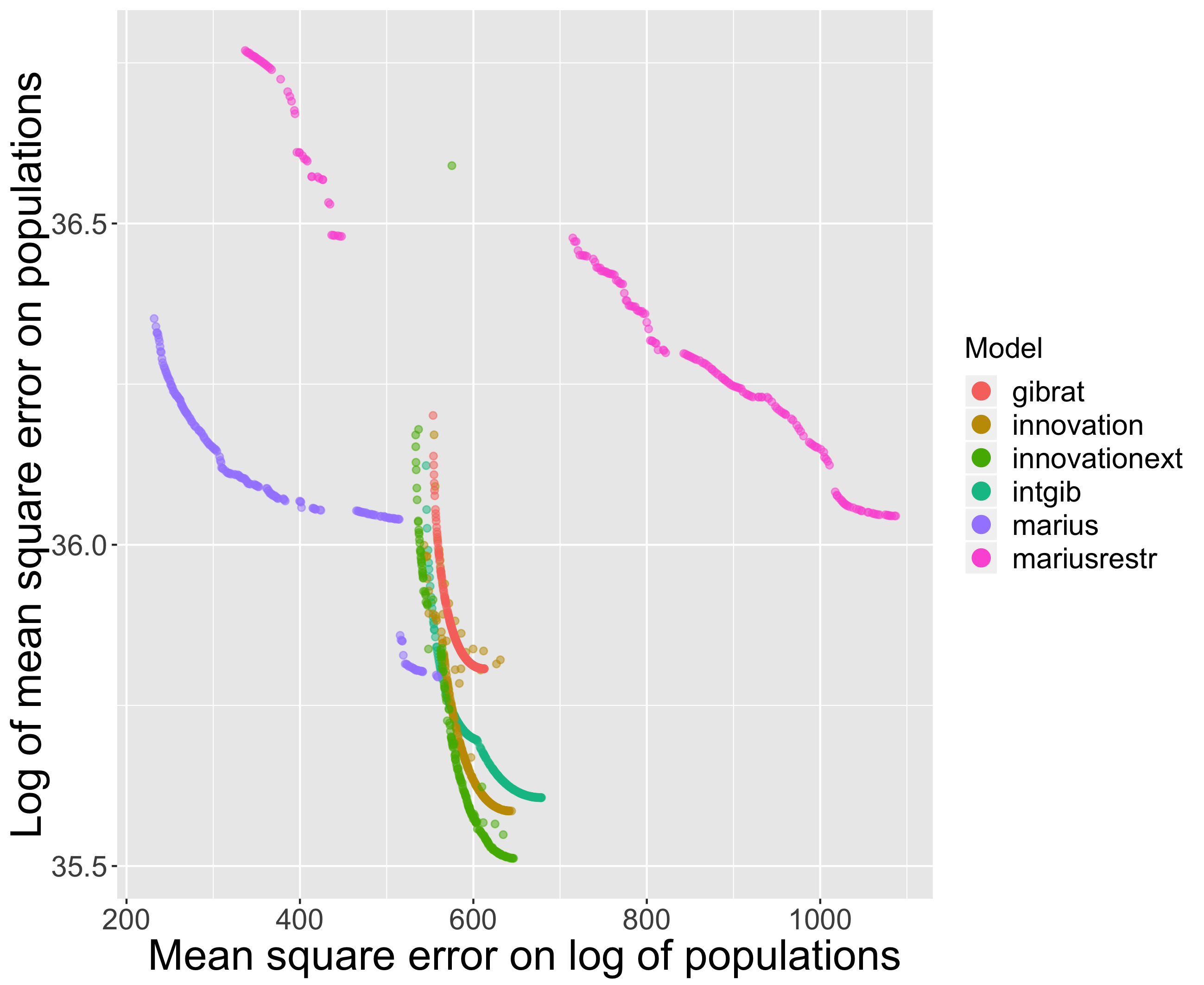}
\caption{Test of six dynamic models simulating urban trajectories on the world system of cities (GHSL data source).\label{fig:fig9}}
\end{figure}

All these models can be formulated within a common framework which has been described by \cite{raimbault2020indirect}. First of all, we consider a deterministic version of the Gibrat's model, for which the extensions with interactions will capture the covariance structure between population trajectories. In the Gibrat's model, formulated as $P(t+1) = R(t) \cdot P(t)$ where $P$ and $R$ are independent random variables, we have $\mathbb{E}\left[P\right](t+1) = \mathbb{E}\left[R\right](t) \cdot \mathbb{E}\left[P\right](t)$. Furthermore, city trajectories are assumed independent, i.e. $\textrm{Cov}\left[P_i,P_j\right] = 0$ for any cities $i \neq j$. Writing $\mathbb{E}\left[P_i\right](t) = \mu_i(t)$, we generalize the deterministic formulation above to a non-linear one by taking $\mu_i(t+1) = f(\mu_i(t))$. The specification of the transition function or algorithm between these average populations will fully determine the model. The corresponding Gibrat's model (named ``gibrat'' on Figures~\ref{fig:fig5} to \ref{fig:fig9}) has one single parameter which is the average endogenous growth rate. Note that in this deterministic version, there is no additional parameter for the variance (or other moments depending on the distribution chosen) of growth rates.

The network interaction model proposed by \cite{raimbault2020indirect} includes linearly a Gibrat's component of fixed endogenous growth rates, a spatial interaction component given by the average interaction potential with all other cities for each city (with Euclidian or network distance, see \cite{raimbault2020indirect} for details) and a second order term of network flow feedback that we do not include here for simplicity. We consider two versions of the model, one (named ``intgib'' for ``Interaction Gibrat'' on Figures~\ref{fig:fig5} to \ref{fig:fig9}) with Euclidian distance between cities to determine interaction potential, the other (named ``intgibphysical'' for ``Interaction Gibrat Physical'' on Figures~\ref{fig:fig5} to \ref{fig:fig9}) with a physical network distance computed as shortest paths with a terrain slope impedance derived from a global Digital Elevation Model. Both models have the same four parameters, namely endogenous growth rate, weight of interactions, hierarchy of interactions, and geographical range of interactions.

The Favaro-Pumain model for the diffusion of innovation \cite{favaro2011gibrat} considers population of cities and additional variables representing an adoption rate of a given innovation. To evolve populations, (i) innovations are diffused in the network of cities following a spatial interaction model and with an intensity depending on the utility of the innovation; (ii) population are updated following another spatial interaction model, interaction potential being driven by the innovative characters of cities; (iii) we introduce exogenously a new innovation with an increased utility if a certain global adoption threshold is reached for the previous innovation, at a fixed initial penetration rate and in a city chosen with a probability calculated according to a scaling law of population. A first simplified version of this model (named ``innovation'' on Figures~\ref{fig:fig5} to \ref{fig:fig9}) has default parameter values from \cite{favaro2011gibrat} and four free parameters which are the endogenous growth rate, the weight of interactions, interaction range for innovation diffusion and interaction range for population growth. The full version (named ``innovationext'' for ``Innovation extended'' on Figures~\ref{fig:fig5} to \ref{fig:fig9}) has nine parameters, with additional parameters being the initial utility of the first innovation, the fixed growth rate of innovation utilities, the initial penetration rate, the adoption threshold for a new innovation, and the hierarchy exponent to determine innovative cities.

Finally, the Marius model family based on economic exchanges \cite{cottineau2014evolution} implements the following processes. Cities are attributed an initial wealth following a scaling law of populations. At each time step, (i) supply and demand are updated for each city as superlinear functions of populations; (ii) cities exchange goods according to a spatial interaction potential and their supply and demand, and wealth are updated accordingly; (iii) population are updated such that population difference follows a scaling law of wealth difference with a given economic multiplier and exponent. A restricted Marius model (named ``mariusrestr'' for ``Marius restricted'' on Figures~\ref{fig:fig5} to \ref{fig:fig9}) has four parameters, namely economic multiplier, supply and demand exponents, and the interaction distance. The full model (named ``marius'') has six parameters, adding the exponent for the initial wealth and the exponent for the population update.

Note that besides the benchmark Gibrat’s model, we have a version of each model with four parameters, which makes them directly comparable in terms of adjustment performance. Indeed, taking into account over fitting in such simulation models remains an open question as \cite{raimbault2020indirect} puts it, and a fair model comparison is ensured with the same number of parameters. We will however consider all model versions in the comparison and study absolute performance of models whatever their number of degrees of freedom. The models are implemented in Scala within a single library, and integrated into the OpenMOLE model exploration platform for exploration and calibration~\cite{reuillon2013openmole}. Source code and results are available on the repository of the project at \url{https://github.com/JusteRaimbault/UrbanGrowth}.

We show in Figure~\ref{fig:fig5} how different dynamic models succeed in simulating the population trajectories of all cities for BRICS countries, Europe and the United States. Here each model is adjusted on the GeoDiverCity dataset for each system of cities, for time spans covering 1960 to 2010 (precise dates depend on each system). Evaluating models on this dataset rather than on the GHSL database seems more relevant for these systems of cities, as data was specifically tailored for comparability and following a consistent geographical definition of urban areas with more refined estimation of their populations \cite{pumain2015multilevel}. The model calibration procedure is the following. Cities are initialized with observed population at the first date in the dataset. Given a model and associated parameter values, populations are then simulated for each date in the dataset according to the model. The fitness is evaluated with two complementary indicators: (i) the logarithm of the total mean square error between observed and simulated populations, taken in time and for all cities; (ii) the mean square error on logarithms of observed and simulated populations taken in time and for all cities. These two indicators are complementary, because of the hierarchical nature of urban systems: considering only the mean square error will favor the adjustment on very large cities only, while considering logarithms of population will put a higher emphasis on the role of medium-sized and small cities, which must not be neglected when considering an urban system \cite{aveline2018pathways,denis2017subaltern}. A multi-objective calibration algorithm (more precisely the NSGA2 algorithm), implemented in the OpenMOLE platform with a specific design to be distributed on a computation grid \cite{schmitt2015half}, was run with these two objectives for each city system and each model. The algorithm is stopped after a fixed number of steps, for which convergence was always obtained (in the sense of negligible variation in the final populations obtained).

The curves of different color in Figure~\ref{fig:fig5} are representing the Pareto fronts, i.e. the points forming an optimization compromise between the two objectives, for each system of cities (subpanels) and each model (colors). The curves represent Pareto fronts obtained with the final generation of the genetic algorithm, corresponding to 200 simulations of each model (one point corresponds to one simulation). On the whole, models with innovation and interaction are performing better for reconstructing the trajectories than the simplest stochastic model, but the patterns of fitness are very diverse. In some cases such as South Africa, Russia and Europe, the full Marius model is clearly dominating all other models, since its Pareto front performs better than the Pareto fronts of other models regarding the two objectives. In these cases, economic exchanges are thus better candidates than the other processes considered to explain urban growth. These are the most mature systems among the ones under study, what would suggest a correspondence between the age of the system and the fact that its dynamics are driven by economic exchanges (furthermore whatever the political and economic system in place, since Europe and Russia before and after the fall of USSR are covered). In some cases such as China, several models are in close competition, while in the remaining cases, different models are complementary to obtain the effective Pareto front.

We give more details on these cases where different models appear to be complementary. Figure~\ref{fig:fig6} and Figure~\ref{fig:fig7} represent contrasted patterns of the Pareto fronts in the case of India and Brazil. In the case of India, for the models shown here Gibrat's model provide the worse predictions of urban trajectories, whatever their position in the urban hierarchy. Obviously introducing interactions, and especially physical interactions, is a necessary and major improvement for simulating the development of Indian cities of all sizes. The interaction model and its physical version (blue and light green) are the best but very close to the full innovation diffusion Favaro-pumain model (dark green). In that case, the two processes are equivalent candidates to explain urban growth. Note that we show here a zoomed region of the previous plots in Figure~\ref{fig:fig5}, and that in that case the Marius economic model is outside the plot range, even dominated by the one-parameter Gibrat’s model. Several explanatory factors for this dominating role of spatial interactions in the Indian system of cities could be proposed, such as a history of a stratified urbanization due to successive colonial extraction periods, but also the capture of a long distance commerce which spans beyond the boundaries of India or its current insertion within globalized value chains.

For Brazil (Figure~\ref{fig:fig7}) the pattern is more complicated because multiple factors differentiate urban trajectories. The Pareto front is formed by the full Favaro-Pumain model (\textit{innovationext}, dark green). The innovation model is always the best whatever the city size considered,  what could be consistent with the historical and geographical context of Brazil: innovations spilling out of the newly founded Brazilia should necessarily have impacted surroundings medium-sized towns. However, we also obtain the physical interaction model (\textit{intgibphysical}, light blue) within the zoomed windows as a second Pareto front. This can be interpreted as also a good explanatory power for spatial interactions taking into account the physical terrain. In that sense, the uneven topography of Brazil must play a role in shaping transportation networks and urban interactions, which is consistent with inequalities in extraction potential, in particular for agriculture.

The case of Chinese cities is interesting, not only because of the peculiarities of the urbanization process in this rapidly urbanizing country with a still very old system of cities since a half century \cite{wu2020emerging}, but also because the complication of local sources of territorial information about urban populations \cite{swerts2013systemes,swerts2017data}. What emerges from Figure~\ref{fig:fig8} that represents the fitness of dynamic models is a striking quasi-equivalence between all models considered, except for the Marius model (purple) slightly dominating the other Pareto fronts when medium-sized cities are privileged. Otherwise, all models are very close to the Gibrat's model (red). Such an ``anomaly'' when compared to the simulations made on all other regions of the world can be interpreted to be produced by the systematic character of the Chinese urban planning aiming at developing urban areas in a parallel way all over the Chinese urban regions. Models introducing spatial interaction produce patterns expected when geographical bottom-up processes of interaction are considered only, but they cannot anticipate highly top-down planned urban development actions.

We have also applied these models to the full set of cities in the world as documented from the GSHL database (Figure~\ref{fig:fig9}). Only such a broad coverage database, despite its potential bias, can allow applying such interaction models at this global scale, and this application thus shows the complementarity of data sources and the potentialities of GHSL. Although we can expect a larger diversity of urban trajectories at that scale because of the much contrasted demographic behavior in regions where the urban transition is achieved and those where it is still going on, this does not seem to hamper the performance of the models. Gibrat's remains the most approximate way of predicting urban trajectories, but what emerges at that world scale is a stronger differentiation between the ability to simulate the top or the bottom of urban hierarchies. Largest cities are rather well approximated with the innovation diffusion model (green), while smaller cities are better adjusted with the Marius economic model. This would be consistent with metropolization processes implying that large global cities interact more between themselves than with their hinterland, at least for highly innovative and value-added activities. Meanwhile, on a regional scale for medium-sized cities (these interacting in smaller ranges than large cities because of the size term in the spatial interaction model), economic exchanges are driving urban growth. 

To what extent such a worldwide estimation is relevant, compared to spatial locally stationary estimations on fixed (as done before) or on variable spatial extents, remains to be investigated. We however show how such models can be applied and compared worldwide, opening research avenues towards systematic benchmarks of dynamical models for systems of cities all over the world. Furthermore, we showed the complementarity of the different models, since none was dominating the others in all cases, confirming the diversity of co-evolution niches suggested before.

\section{Towards multi-scale models}

While several applications and developments of the studies described previously will be necessary to strengthen our knowledge of evolutive urban systems, the development of new kinds of models will also be a crucial feature towards sustainable territorial governance. Following \cite{pumain2008socio}, urban systems imply diverse processes at different scales, with upward and downward feedbacks between these. The typical scales are first the intra-urban scale (microscopic scale), implying urban stakeholders, built environment, physical artifacts; second the metropolitan scale, or mesoscopic scale, for which the important processes are the location of population, economic activities and amenities, and for which abstractions done for example in Land-use Transport (LUTI) models are good approximations to grasp urban dynamics \cite{wegener2004land}; and third the scale of the system of cities, or the macroscopic scale, for which spatial interactions and the processes described in the model we used are appropriate abstractions. This view considerably extends and details the formulation of \cite{berry1964cities} of cities as systems of systems. In that context, \cite{rozenblat2018conclusion} point out that a multiscalar methodology is necessary for territorial policies, for example to be able to articulate global issues with local issues in the least contradictory way possible, with multiple and generally conflicting objectives. \cite{raimbault2019multi} illustrates this by empirically studying the economic and ecological performance of European mega-urban regions with a multi-level perspective through endogenous definitions of these. Furthermore, the recent transition from relatively modular regional urban systems to a globally interconnected urban system, and possibly associated new settlement patterns such as polycentric mega-urban regions \cite{le2017etalement}, implies a need for new models to possibly understand how this transition occurs, or simulate urban dynamics after the coalescence of regional urban systems.

An example of such multi-scalar models can be given between the mesoscopic and macroscopic scale. The macroscopic dynamics of a metropolitan area will strongly drive its internal development, for example because of population and economic flows coming from outside. This implies a downward feedback from the macroscopic urban dynamics to the mesoscopic dynamics. Reciprocally, the internal organization of a metropolitan area will influence its dynamics and insertion in the system at the macroscopic scale (whatever the dimension, for example economic performance), through positive and negative externalities such as congestion. This consists of upward feedback. A model strongly coupling these two scales and including explicitly these two feedbacks, coupling a reaction-diffusion model of urban growth at the mesoscopic scale \cite{raimbault2018calibration} with the interaction model of \cite{raimbault2020indirect} at the macroscopic scale, has been proposed by \cite{raimbault2019multiscalar}. Model exploration on synthetic systems unveils non-trivial non-linear effects from including the feedbacks, and for example intermediate optimal ranges for policy parameters influencing the level of sprawl (transit-oriented development) or the level of local aggregation (metropolization). The development of such models and their calibration on real dataset such as the GHSL dataset we used here, can become precious tools for evidence-based urban policies.

\section{Conclusion}

The essential point of evolutionary theory is to take into consideration the spatio-temporal dimension of the urban realm The aim is to link the development of cities to the many and diverse interrelations that make cities, since their emergence, entities that are not isolated, but on the contrary interdependent in their evolution, to the point of constituting ``systems of cities''. These systems are social adapters (in the sense that they carry and induce social change), complex, multi-scalar and open. The dynamics of these systems of cities, although they must always be placed in a context of time and space, include regularities that make it partly comparable and predictable, from one system to another and for certain scales of time. It is the micro-geographical level interactions, formed by the multiple interventions of a large number of stakeholders that produce the "behaviors" of cities and city systems at meso- and macro-geographical scales, because of the complex reflexive feedbacks introduced by the practices of so diverse stakeholders. It is important for these people and institutions to be informed of such knowledge about urban dynamics, to take advantage of this collective territorial intelligence and to make the most of the important adaptations required by the ecological and social tensions of our time. For complex dynamics such as the one observed in urban systems, policy interventions are always difficult as they can yield unexpected and even contrary effects. The compromise between an external top-down shock and endogenous bottom-up measures is also a subtle aspect to be determined. Understanding urban dynamics and co-evolution regimes in their diversity brings indirect elements of answer to these policy issues.

What modelling of urban dynamics brings in the discussion is at first a solid confirmation about the robustness of stylized facts that are integrated within the evolutive theory of urban systems. While also confirming the complementarity of processes and models it underlines the importance of the historical/political/geographical context which produces numerous effects of path-dependency.

Many questions remain open, as how to link urban scaling and dynamic models, how to define endogenously consistent urban systems, and how to develop data-driven multiscale models. However, a few messages can be conveyed towards citizens and practitioners: (i) there is statistical predictability of city growth and size on short time periods; (ii) largest metropolises are not “monstruopolises” but “normal” products of the urbanization process in their particular territorial conditions; iii) proactive adaptive strategies are necessary (through imitation, or anticipation and risk) for maintaining every city updating within a context that remain too often conceived as a rivalry or competition but that should more and more evolve towards emulation, according to a concept of co-opetition. We can have confidence in the future of cities because of the well observed robustness and sustainability of systems of cities, and the wide variations in their organization and evolution remind us about the fact that there are no norms nor any optimum in the territorial, social and cultural design of cities. Their diversity demonstrates the viability of different ways of being urban, and perhaps is a guarantee of the sustainability of the systems that they construct at the world scale.




\section*{Acknowledgments}

Results obtained in this paper were computed on the vo.complex-system.eu virtual organization of the European Grid Infrastructure (http://www.egi.eu). We thank the European Grid Infrastructure and its supporting National Grid Initiatives (France-Grilles in particular) for providing the technical support and infrastructure.


\end{document}